\documentclass{article}
\RequirePackage[a4paper]{geometry}
\usepackage{amsmath,graphicx}

\begin{document}
\renewcommand{\abstractname}{\vspace{-\baselineskip}}

\begin{center}
{\Large\bfseries
Photonic reservoir computing based on nonlinear wave dynamics at a
 microscale
\par}
\vspace{3ex}
{\bfseries
Satoshi Sunada$^1$ and Atsushi Uchida$^2$
\par}
{\footnotesize\itshape
$^1$Faculty of Mechanical Engineering, Institute of Science and
Engineering, Kanazawa University\\
Kakuma-machi Kanazawa, Ishikawa 920-1192, Japan \\
$^2$Department of Information and Computer Sciences, Saitama University,\\
255 Shimo-Okubo, Sakura-ku, Saitama City, Saitama, 338-8570, Japan.\\
\par}
\vspace{3ex}
\end{center}

\begin{abstract}
High-dimensional nonlinear dynamical systems including neural networks
 can be utilized as a computational resource for information
 processing. 
In this sense, nonlinear wave systems are good candidate for such a
computational resource. 
Here, we propose and numerically demonstrate information processing based on nonlinear wave dynamics in
 microcavity lasers, i.e., optical spatiotemporal systems at a microscale.   
One of the remarkable features is the ability of high-dimensional and nonlinear mapping of input
 information into the wave states, enabling efficient and fast
 information processing at a microscale. 
We show that the computational capability for nonlinear/memory tasks is
 maximized at the edge of the dynamical stability.
Moreover, we also show that the computational capability can be enhanced 
by applying a time division multiplexing technique to the wave dynamics; thus, 
the computational potential of the wave dynamics can sufficiently be extracted
 even when the number of detectors to monitor the wave states is limited.
In addition, we discuss the merging of optical information processing 
and optical sensing,
opening a novel method of model-free sensing by using a microcavity reservoir itself as a sensing element.
These results open a way to on-chip photonic computing with
 high-dimensional dynamics and a novel model-free sensing scheme. 
\end{abstract}

\section*{Introduction}
Reservoir computing (RC) \cite{Versraeten2007}, originally known as an
echo state network \cite{Jaeger2004} or a liquid state machine
\cite{Maass2002}, is a computational
paradigm using high-dimensional dynamical systems, and 
it has been regarded as a powerful tool for solving highly-complex and
abstract computational tasks.
The computational paradigm has recently been implemented in a variety of
physical systems or devices, such as optoelectronic systems
\cite{Paquot2012}, photonic systems \cite{GVderSande2017}, memristors
\cite{Du2017}, spin systems \cite{Nakane2018}, and soft materials
\cite{Nakajima2015}. 
(See Ref.~\cite{Tanaka2019} for a comprehensive review on physical RCs.)
In particular, the photonic implementation of RC has been expected to open 
the path to ultrafast and efficient processing beyond traditional
Turing-von Neumann computer 
architecture \cite{Brunner2013,Larger2017,Vandoorne2014,Takano2018,Laporte2018,Sunada2018}. 

A key principle of the RC is a high-dimensional mapping of the
input information based on the high-dimensionality of the reservoir dynamical
systems;
the computational capability is dependent on the number of linearly
independent internal states of dynamical systems in response to an
encoded input \cite{Dambre2012}. 
Moreover, a nonlinearity and short-term memory effect inherent in
dynamical systems also play a crucial role in solving computational tasks requiring
nonlinearity or memories.
Thus, infinite dimensional nonlinear systems will be good candidates as reservoirs.

One of the representative infinite dimensional systems 
is a delay system, where 
reservoir networks can virtually be constructed in a time domain \cite{Appeltant2011}.  
To date, numerous experimental studies in the photonic RC
with delay systems have been performed
because of the easy implementation in optoelectronic or photonic
systems, such as lasers
with delayed feedback \cite{Brunner2013,Larger2017,Martinenghi2012,Duport2012}.
The information
processing, including prediction \cite{Brunner2013} and speech
recognition \cite{Larger2017}, have been
demonstrated;
however, the drawback is the requirement of long delay lines to make many
virtual nodes in the RC, which could lead to impractically large
systems, inhibit stable operation , and may prevent practical
deployments.   

In this study, we propose to use a microcavity laser, i.e., 
a microscale spatially-extended optical system, as a reservoir. 
Originally, microcavities have mainly been utilized to realize a low threshold laser
source and modify quantum effects by the strong optical confinement effect, which is caused by the
difference in refractive indexes between inside and outside the cavity
\cite{OpticalMicrocavity1996}.  
Then, a various shapes of microcavity lasers, inspired by wave/quantum
chaos, have recently been utilized to control the emission
properties \cite{NS1997,Cao2015}. 
An interesting feature of such microcavity lasers is to be able to 
exhibit a variety of spatiotemporal wave dynamics by the interplay of a
gain medium and cavity shape \cite{Harayama2011,Bitter2018}.
Unlike the previous works, we utilize such wave dynamics in microcavity 
lasers driven by an input signal for RCs and
numerically demonstrate that 
the RC-based information processing 
can efficiently be achieved {\it at a microscale} owing to the spatial
degrees of freedom based on the high-dimensional dynamics with a long memory effect. 

In addition, we discuss the application of the microcavity-based
processing by using the sensitivity of wave dynamics in a microcavity 
to an external perturbation; 
we propose to use a microcavity as a sensing element as well as a
reservoir, resulting in high-dimensional mapping. 
The merging of optical sensing and the reservoir suggests the possibility of novel
 sensing without complex post processing and theoretical sensing
models. 
As a proof-of-concept demonstration, we show fast sensing of external reflective index by using the microcavity RC.  

\section*{Microcavity-based RC}
\begin{figure}
\begin{center}
\raisebox{0.0cm}{\includegraphics[width=14cm]{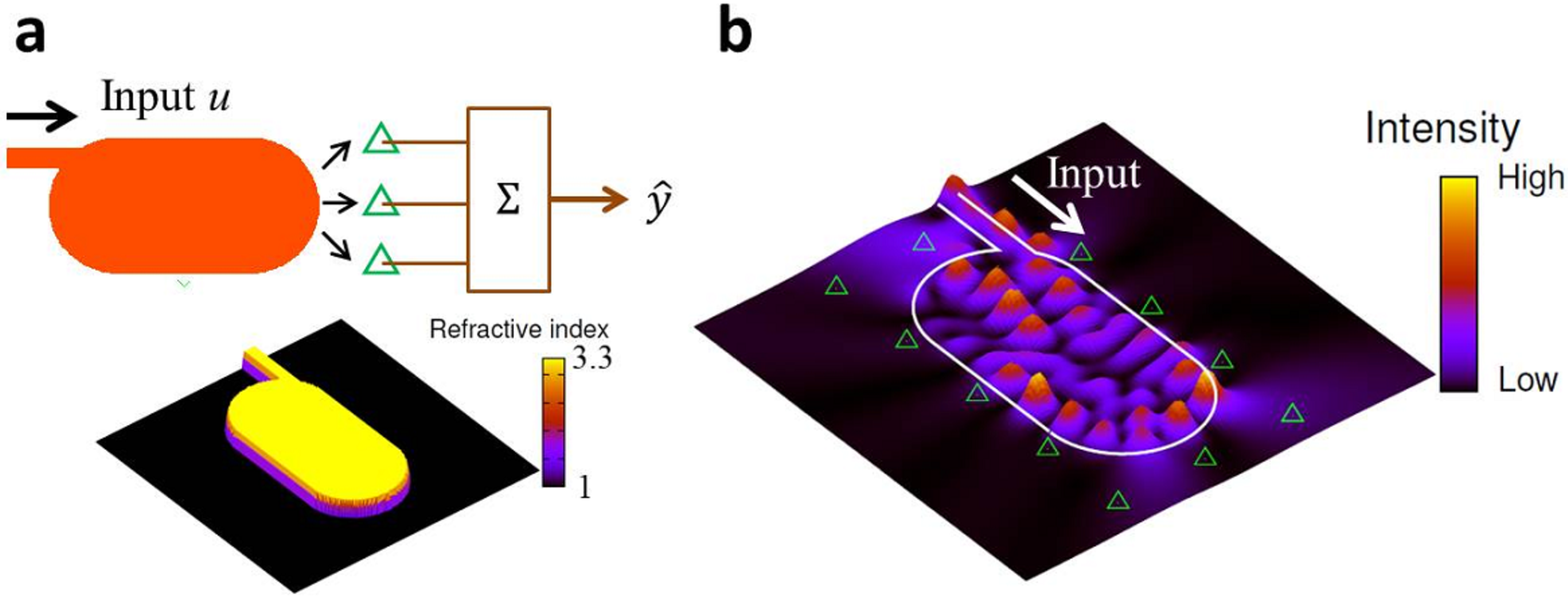}} 
\end{center}
\caption{\label{fig1} 
{\bf Microcavity laser for RC.}
{\bf a.} Schematic of RC using a microcavity laser.
The cavity shape is designed as the Bunimovich's stadium, 
known as a chaotic cavity and the cavity is coupled to an input
 waveguide. 
The incident light encoded by the signal $u$ is injected to the cavity, 
and the emitted light intensities are detected by the probes, represented by the green
 triangles, and they are used for the output $\hat{y}$. 
The lower figure in {\bf a} shows the refractive index distribution of
 the stadium cavity coupled to an input waveguide.
$n_{in}=3.3$ and $n_{out}=1.0$ denote the refractive indeces inside and outside
 the cavity and waveguide, respectively. 
{\bf b.} An example of the wave patterns responding to the input light, 
the frequency of which is tuned to a resonant frequency of the cavity $\omega_0$.  
The boundary of the microcavity coupled to the waveguide is 
represented by the white curves. 
The green triangles represent the point probes to detect the emission 
from the microcavity lasers.
In the simulation, the normalized pumping power $W_{\infty}/W_{th}\approx 1$ was
 set, where $W_{th}$ is the threshold pumping power. 
}
\end{figure}

\begin{figure}
\begin{center}
\raisebox{0.0cm}{\includegraphics[width=12cm]{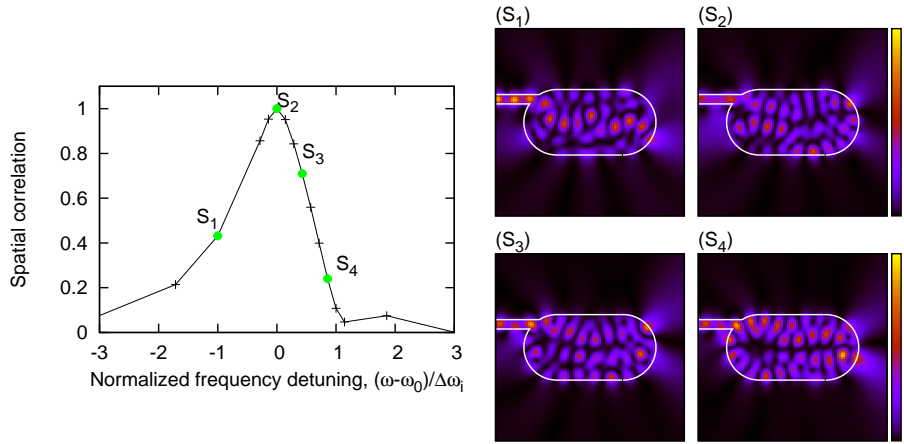}} 
\end{center}
\caption{\label{fig2} 
{\bf Dependence of excited wave patterns on the input frequency $\omega$}.
The left figure shows the spatial correlation 
$C(\omega)=\langle
 (I_{\omega}-\bar{I}_{\omega})(I_{\omega_0}-\bar{I}_{\omega_0})\rangle /(\sigma_\omega\sigma_{\omega_0})$, 
where 
$\langle\cdot\rangle$ denotes the spatial average, 
and $I_{\omega}(\mbox{\bf r})$ represents the intensity pattern excited by the input
 light with frequency $\omega$. 
$\bar{I}_{\omega(\omega_0)}$ and $\sigma_{\omega(\omega_0)}$ denotes the
 spatial average and standard deviation. 
The horizontal axis is the frequency detuning of the input light 
from a resonant frequency $\omega_0$ of the stadium cavity, 
normalized by the average mode interval $\Delta\omega_i$. 
As seen in this figure, the input frequency affects the wave patterns; thus
the input information can be encoded into the wave patterns with the
 instant frequency of the phase-modulated light. 
In the simulation, $W_{\infty}=0$ was set. 
}
\end{figure}

Figure \ref{fig1}(a) shows a schematic of the proposed system, which
consists of a microcavity coupled to an input waveguide and probes
(detectors) to make the RC output $\hat{y}$.   
The microcavity include a nonlinear gain medium, and 
the cavity shape is designed as the Bunimovich stadium \cite{Stadium}, in which 
ray orbits are proven to be fully chaotic and the corresponding wave patterns are
complex (Fig. \ref{fig1}b).
An optical signal encoded with a phase modulation is injected from the
input waveguide and can reach all parts of the cavity due to the chaotic
multiple reflections at the cavity boundary and is nonlinearly amplified
by the gain medium. 
A feature of the stadium cavity is the dependence of the wave
 pattern on the input frequency; the wave pattern sensitively changes,
depending on the input frequency.
Actually, as demonstrated in Fig. \ref{fig2}, the spatial correlation between 
two wave patterns excited by the inputs with frequencies $\omega_0$ and
$\omega=\omega_0+\Delta\omega$ decreases as $\Delta\omega$
increases. 
This means that 
the information can be encoded into the wave patterns
with the instant frequency by phase-modulating the input light.
Then, the gain medium play an important role in adding an additional
nonlinearity and memory effect by the amplification. 
The emitted signals from the cavity are detected at point probes 
at the sampling time interval $\tau_s$.
In the simulation, $N$ probes are assumed to be placed around the
cavity.  

For RC, we consider the linear readout $\hat{y}$ as $\hat{y}(t)=\sum_{i=1}^Mw_ix_i(t)$, where 
$x_i$ is the detected intensity at probe $i$, ($i\in\{1,2,\cdots,N\}$),
at time $t=n\tau_s$
($n\in \{1,2\cdots, \infty \}$), and $w_i$ is a readout weight. 
The goal of the processing is to approximate a functional
relation between input signal $u(n)$ and target signal $y(n)$ by the
readout $\hat{y}$. 
To this end, a finite set of training data $\{u(n),y(n)\}_{n=0}^T$ is
utilized to determine the readout weights, such that the normalized mean
square error $1/T\sum_n|y(n)-\hat{y}(n)|^2$ is minimized.   
In the training process, we simply use the least square method. 

\section*{Results and discussions}
To gain an insight into the computational capability of the
microcavity-based RC, the numerical simulation was performed by using the Maxwell-Bloch (MB)
model, where the gain medium is modeled as a simple two-level system \cite{Harayama2005}. 
Whereas the MB model is a simple model of microcavity
lasers, the dynamical lasing phenomena can qualitatively be examined
\cite{Harayama2003,Sunada2013}. 
We assumed that the cavity is two-dimensionally extended on a plane, and 
the electric field is polarized perpendicular to the plane.
For generality, all variables were made dimensionless (see {\it Methods}
for details), 
and we discuss the RC capability with the dimensionless
variables. 

\begin{figure}
\begin{center}
\raisebox{0.0cm}{\includegraphics[width=12cm]{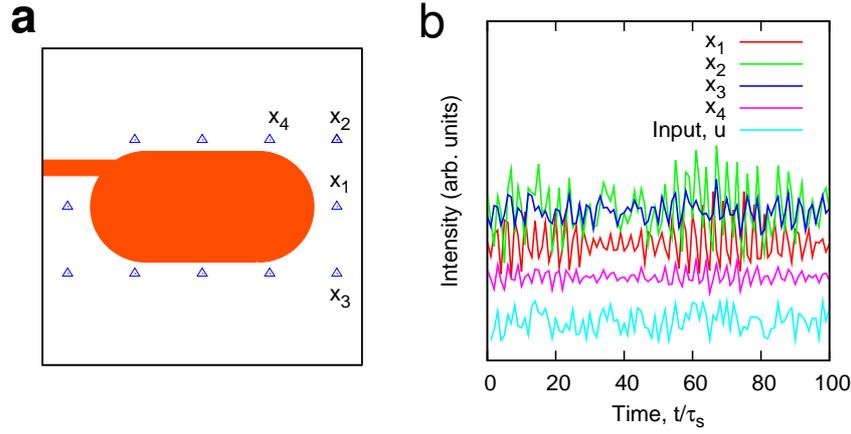}} 
\end{center}
\caption{\label{fig3} 
{\bf Demonstration of emission dynamics.}
{\bf a.} 
A schematic of the stadium-shaped microcavity laser coupled to an
 input waveguide. 
The blue triangles denote the positions of the point probes. 
{\bf b.} Intensity signals $x_i$ detected by the probes $i=1,2,3$ and 4 
when the input light is phase-modulated with a signal $u$.
}
\end{figure}

In the simulations, the refractive index $n_{in}$ inside the
cavity was set to be 3.3, the length $L$ of the major axis of the stadium cavity was
$\approx$1.67$\lambda$, where $\lambda$ is the wavelength of 
the input light in vacuum. 
(If $\lambda =0.85$ $\mu$m, $L$ would be 1.42 $\mu$m.) 
$\tau_s$ was fixed to be close the lifetime
$\tau_{c}$ of the cavity without a gain medium.
(See {\it Methods} for discussions toward actual experiments.)
The input information $u(n)$ was encoded in the phase of the input light as
$\phi(t) = m_a u(n)$, where $m_a \approx 0.1$ is the modulation amplitude. $u(n)$
holds for the time interval $\tau_s$.
The center frequency of the input light was locked to a resonant 
frequency $\omega_0$. 
Under these conditions, a variety of the intensity signals were measured 
in response to the modulated signal, as demonstrated in
Fig. \ref{fig3}. 

\subsection*{Nonlinear-memory capacity.}
\begin{figure}
\begin{center}
\raisebox{0.0cm}{\includegraphics[width=12cm]{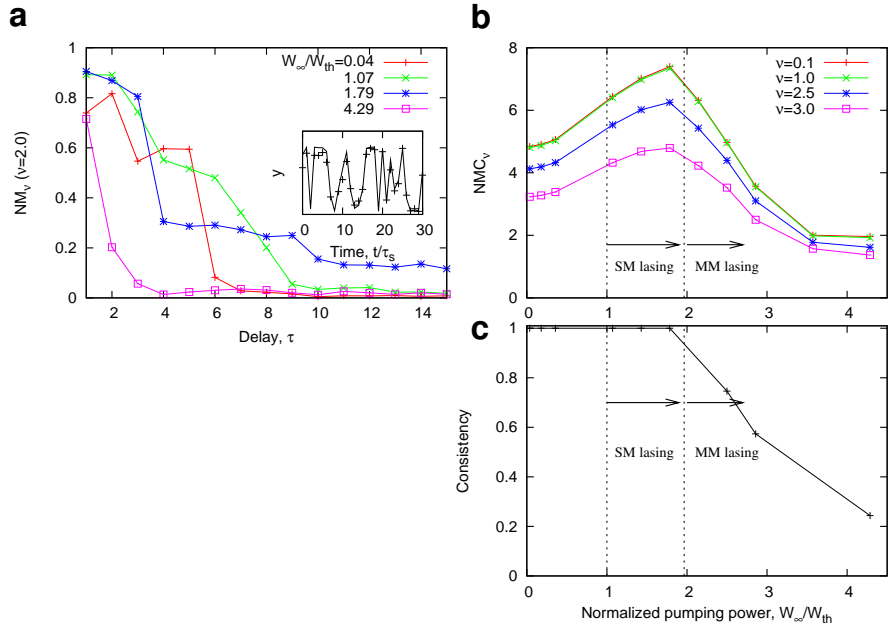}} 
\end{center}
\caption{\label{fig4} 
{\bf Nonlinear memory capacity of a stadium-shaped microcavity laser.}
{\bf a.} $NM_{\nu}(\tau)$ for $\nu=2.0$. $W_{\infty}/W_{th}$
 denotes the pumping power normalized by the threshold pumping power. 
In the simulation, $N=11$ probes were used. 
The inset shows an example of the RC output $\hat{y}(n)$ for
 $W_{\infty}/W_{th}\approx 1.79$, denoted by the
 crosses, and
 the target signal $y(n)$ for $\nu=2$ and $\tau=1$ (solid black curve) in time $t/\tau_s$. 
{\bf b.} $NMC_{\nu}$ as a function of the normalized pumping power $W_{\infty}/W_{th}$. 
When $W_{\infty}/W_{th}>1$, a single mode (SM) lasing starts, and
 multimode (MM) lasing occurs when $W_{\infty}/W_{th}>1.79$.
For the large pumping power $W_{\infty}$, the loss of the input
information is compensated and a long memory effect can be achieved. 
In addition, nonlinear gain saturation plays an important role in
introducing additional nonlinearity in the reservoir.
However, consistency decreases when MM lasing occurs. 
Consequently, $NMC_{\nu}$ is maximized around the edge of the stability.  
{\bf c.} Consistency as a function of $W_{\infty}/W_{th}$. 
See {\it Methods} for the measurement method of the consistency. 
}
\end{figure}

For the evaluation of the computational capability, 
we use a simple function approximation tasks, $y(n)=\sin[\nu
u(n-\tau)]/\nu$, where $\nu$ and $\tau$ are the task parameters that control
the required nonlinearity and memory, respectively \cite{Inubushi2017}.
The input signal $u(n)$ is an identically
distributed random sequence generated from a uniform
distribution between $[-1,1]$.
The goal of the task is to reproduce the nonlinearly-converted signal
$y(n)$ with a delay of $\tau$ (see the inset of Fig. \ref{fig4}a). 
To evaluate both the ability to adapt nonlinear tasks and memory
capacity of the RC, 
we introduce the correlation between the target signal $y$ and output $\hat{y}$,
\begin{eqnarray}
NM_\nu(\tau)
= 
\dfrac{\langle y(n-\tau)\hat{y}(n)\rangle^2}
{
\sigma^2_y\sigma_{\hat{y}}^2
},
\end{eqnarray} 
where $\langle\cdot\rangle$ is the mean over time step $n$,
$\sigma_{z}$ denotes the standard deviation of $z=y$ or $\hat{y}$.
Then, the nonlinear-memory capacity is defined as the sum of
$NM_{\nu}(\tau)$, with $\tau$ going to infinity:
\begin{eqnarray}
NMC_{\nu} = \sum_{\tau=0}^{\infty}NM_{\nu}(\tau).
\end{eqnarray}
$NMC_{\nu}$ corresponds to the linear memory capacity $MC$ in the limit
of $\nu\rightarrow 0$ \cite{Ortin2015}. 
With the nonlinear-memory capacity $NMC_{\nu\ne 0}$, 
we can both evaluate the nonlinearity and memory effects in microcavity
lasers at the same time.  

We examined the lasing dynamics in the stadium cavity and obtained $NMC_{\nu}$ from the 
intensity signals detected by each probe.
Figure \ref{fig4} shows the numerical results of $NM_{\nu}(\tau)$ and
$NMC_{\nu}$ with various values of the pumping power $W_{\infty}$ in the
gain medium (see {\it Methods} for $W_{\infty}$). 
As shown in Fig. \ref{fig4}a,
$NM_{\nu}$ decreases with increasing $\tau$, but 
the decrease becomes moderate by increasing $W_{\infty}$ 
in a range of $0\le W_{\infty} \le 1.79W_{th}$,
where $W_{th}$ denotes the threshold pumping power. 
The pumping compensates the loss of the input information by the nonlinear
amplification effect in the gain medium, and the reservoir (cavity) 
can have a longer memory and becomes adaptive to nonlinear tasks.  
Accordingly, $NM_{\nu}$ and
the resulting $NMC_{\nu}$ increases with increasing $W_{\infty}$ (Fig. \ref{fig4}b).
However, when $W_{\infty} > 1.79W_{th}$, multimode lasing occurs
and $NMC_{\nu}$ decreases.
Consequently, $NMC_{\nu}$ of the RC with the microcavity laser is
maximized around the edge of the phase transition, 
$W_{\infty}/W_{th} \approx 1.79$. 

The decrease of $NMC_{\nu}$ is explained by the loss of the consistency \cite{Uchida2004}, or
 the so-called echo-state property \cite{Jaeger2004}, which is an important condition for
 RC \cite{Inubushi2017,Nakayama2016}.  
In the multimode lasing regime, the spontaneous multi-modal oscillations appear, 
leading to different results even from the same input, depending on the
initial states of the reservoir; thus, the appearance of the
irreproducibility prevents consistent processing of the input information. 
We measured the consistency, which is defined by the mean correlation between
the output signals starting from two different initial states. 
(see {\it Method} for the detail.)  
As shown in Fig. \ref{fig4}c, the decrease of the consistency is linked
to the degrade of $NMC_{\nu}$. 

As shown in these results, 
the effects of the gain medium as well as the high-dimensionality of the wave states 
play a crucial role in enhancing the computational capability in the RC
frameworks. 
We emphasize that the compensation of the short memory inherent in compact RC systems 
and additional nonlinearity caused by the interaction with the gain medium 
are advantage, compared to conventional {\it passive} photonic 
integrated RC \cite{Vandoorne2014,Laporte2018,Sunada2018}, 
where nonlinearity is introduced only in the measurement process. 

\subsection*{Effect of cavity shapes.}
\begin{figure}
\begin{center}
\raisebox{0.0cm}{\includegraphics[width=12cm]{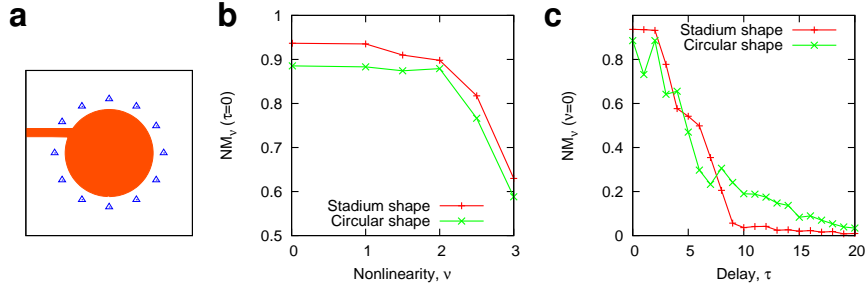}} 
\end{center}
\caption{\label{fig5} 
{\bf Performance comparison of stadium- and circular-shaped lasers.}
{\bf a.} A schematic of the circular-shaped laser coupled to an
 input waveguide. 
The blue triangles denote the positions of the point probes. 
{\bf b.} Adaptivity to nonlinear tasks, where $NM_{\nu}$ with $\tau=0$
 is shown in a function of the nonlinear parameter $\nu$.   
{\bf c.} Linear memory capacity. $NM_{\nu=0}(\tau)$ is shown as a
 function of the delay parameter $\tau$.
In {\bf b} and {\bf c}, the performances of the stadium- and
 circular-shaped lasers are compared under the same input
 condition, same number of probes $N=11$, and same pumping condition in
 a single-mode lasing (consistency) regime. 
For nonlinear tasks, the $NM_{\nu}$ of the stadium-shaped laser is
higher than that of the circular-shaped laser, as shown in {\bf b}. 
However, the circular-shaped laser outperforms the stadium-shaped laser 
for tasks requiring long memory $\tau \gg 1$, as shown in {\bf c}. 
}
\end{figure}

The cavity shapes play a crucial role in the quality of the light confinement and 
the wave dynamics, resulting from the multiple reflections inside the cavity.
In the stadium cavity, the chaotic multiple reflections will lead to
efficient wave mixing dynamics, 
enabling high-dimensional mapping of the input
information into complex wave patterns, as demonstrated in the previous subsection. 
To gain a further insight into the effect of the wave-chaotic cavity on 
the RC performance, we also numerically examined the laser dynamics in 
a {\it non-chaotic cavity}, where the internal ray orbits do not exhibit
chaos.
As a non-chaotic cavity, we choose a circular-shaped cavity (Fig. \ref{fig5}a)
and compared the RC performance with that obtained in the stadium-shaped lasers with the same area and same pumping power condition in a
consistency regime.

Figure {\ref{fig5}}b shows the performance comparison for nonlinear tasks, where $NM_{\nu}$ 
is shown as a function of nonlinear parameter $\nu$ when delay $\tau=0$.
Clearly, 
$NM_{\nu}$ of the stadium-shaped laser can outperform that of the circular-shaped
laser for all values of $\nu$, which may partly be attributed to a strong wave mixing
effect in the stadium cavity.
As shown in Fig. \ref{fig5}c, however, for the tasks requiring memory
with the delay parameter $\tau \gg 1$, 
the $NM_{\nu}$ of the circular-shaped laser is relatively higher than that of 
the stadium-shaped laser because the circular cavity 
has longer cavity-lifetime, (lower loss rate) \cite{Cao2015}. 
These results suggests the trade-off between the cavity shapes exhibiting
a long-memory effect and nonlinearity. 
In terms of ray-wave correspondence \cite{Harayama2015}, 
the effect of the cavity shape
becomes more dramatic for a larger value of
the size parameter defined by $n_{in}L/\lambda$; thus, it is expected to 
lead to a larger difference in the RC performance.  
The investigation along this line will be an important issue in the RC. 

\subsection*{Enhancing the computational capability.}
\begin{figure}
\begin{center}
\raisebox{0.0cm}{\includegraphics[width=6cm]{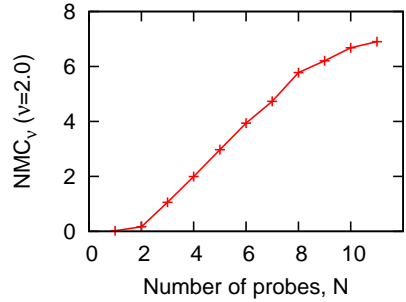}} 
\end{center}
\caption{\label{fig6} 
{\bf $N$-dependence of RC performance.}
$NMC_{\nu}$ vs. the number of probes $N$, corresponding to the
 number of the reservoir nodes.
In the simulation, the minimum distance
 between the probes is set to be close to the wavelength inside the
 cavity, so as to obtain independent responses to the input signals.  
$NMC_{\nu}$ increases as $N$ increases.
}
\end{figure}

In the RC, the computational capability generally increases 
as the number of independent reservoir nodes increases because 
the expressivity increases \cite{Dambre2012}.
In our case, the number of the reservoir nodes used for the output $\hat{y}$ 
correspond to the number $N$ of the signals detected by the probes.
As clearly depicted in Fig. \ref{fig6}, 
 the $NMC_{\nu}$ actually turned out to be proportional to 
the number of the probes $N$. 
This may be one of the merit of using the high-dimensional wave dynamics
in the sense that the capability is enhanced by increasing the number of the probes;
however, the number $N$ 
will be limited by the cavity size and wavelength
because the similar output are obtained 
if the minimum distance between the probes is much shorter
than the wavelength.   
We roughly estimate the maximum number of the probes (effectively corresponding to the maximum number of the measurable nodes) as the ratio of the perimeter $P=(\pi/2+1)L$ of the stadium cavity 
to characteristic wavelength $\lambda/n_{in}$ inside the cavity with $n_{in}$, 
$N \sim (\pi/2+1)n_{in}L/\lambda$, considering that the spatial autocorrelation of the wave patterns typically is sufficiently small for a spatial scale larger than $\lambda$.  
For example, $N\approx 1600$ nodes can potentially be used in 0.013
mm$^2$ footprint for $L=160 \mu$m, $\lambda=0.85$ nm, $n_{in}=3.3$. 
This suggests that larger sized cavity will have a larger computational
capability.   
We emphasize that the potential of implementing such high-density and
large-scale (virtual) nodes is a unique characteristic of the wave
dynamical RC, which is different from conventional photonic integrated RC \cite{Vandoorne2014}, 
consisting of multiple elements.

\subsection*{Using spatiotemporal dynamics for RC.}
As discussed in the previous subsection, the computational capability
depends on the number of the probes $N$ in the present RC.  
In the actual implementation, however, it may be practically difficult to place
a large number of probes (or detectors) around the cavity. 
To overcome the problem, it should be noted that the dynamical
information is included in a delayed sequences
obtained from a few observables \cite{Takens1981}. 
This suggest the possibility that even when only a few
observables are utilized, 
the dynamical information can be extracted from the dynamics of a
few observables. 
In addition, the use of mask signals can make a rich variety of the
reservoir responses \cite{Appeltant2011}.  
We use virtual nodes in a time domain for RC with
a time-multiplexing method used in delay-based RCs
\cite{Appeltant2011,Ortin2015}.  
First, an input signal is multiplied by a mask
signal with a period of $T_m$. 
Then, each response $x_i$ to the signal at probes $i$ is sampled at
$M$ times with a sampling interval $\tau_s(=T_m/M \approx \tau_c)$.
We describe the response $x_i$ at time $t=nT_m+j\tau_s$ 
 $(j\in [1,2,\cdots, M])$ as the node labeled by $i$ and $j$, i.e.,
 $x_{ij}(n)=x_i(nT_m+j\tau_s)$.
Moreover, we use the past node response $x_{ij}(n-k)$, ($k\in \{1,2,\cdots,K\}$)
Finally, the output signal $\hat{y}(n)$ at time step $n$ is calculated as
\begin{equation}
y(n)=\sum_{i=1}^N\sum_{j=1}^M\sum_{k=0}^K w_{ijk}x_{ij}(n-k),
\end{equation} 
where $w_{ijk}$ is an optimal weight, which
can be obtained by using the least-squares method. 

\begin{figure}
\begin{center}
\raisebox{0.0cm}{\includegraphics[width=18cm]{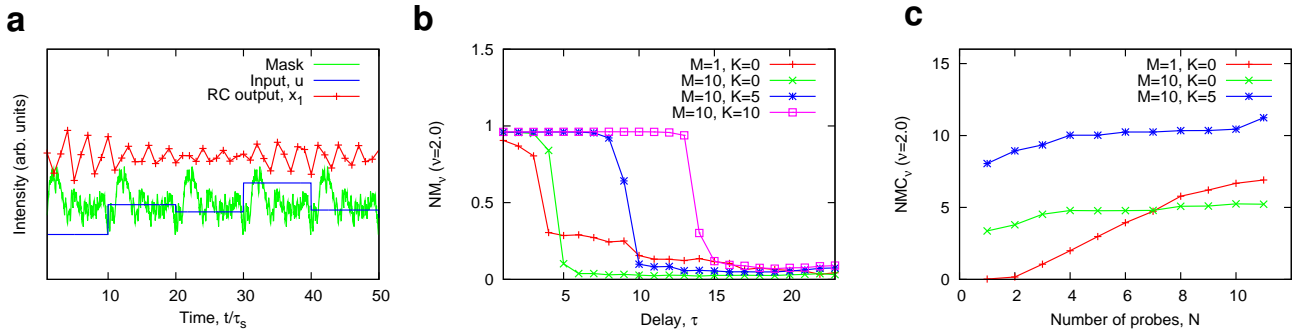}} 
\end{center}
\caption{\label{fig7} 
{\bf Information processing based on the time-multiplexing method.}
{\bf a.} An example of time-multiplexing encoding for $M=10$, where 
a colored random signal with period $T_m$ is used as the mask signal. 
A variety of reservoir responses (e.g., $x_1$) are obtained by the mask signal and
 input $u$.
{\bf b.} $NM_{\nu}$ as a function of the delay parameter $\tau$. 
The $NM_{\nu}$ is enhanced by increasing both $M$ and $K$. 
{\bf c.} $NMC_{\nu}$ vs. the number of probes $N$, corresponding to the
 number of the reservoir nodes. 
$NMC_{\nu}$ can increase as $N$, $M$, and $K$ increase.
Consequently, the large $NMC_{\nu}$ is achieved by the time-multiplexing method 
even when the number of probes is limited.  
In {\bf a.}-{\bf c.}, $W_{\infty}/W_{th}=1.43$. In {\bf a.} and {\bf b.}, $N=11$. 
}
\end{figure}

An example of the time-multiplexing method for $M=10$ is shown in
Fig. \ref{fig7}a, where the input information $u$ holds for the period
$T_m$, and the colored random signals with the period $T_m$ 
is used as the mask signal because  
the use of colored noise or chaotic oscillation as the mask signals
will lead to a good RC performance \cite{Kuriki2018}. 
Figures \ref{fig7}b and c show the $NM_{\nu}(\tau)$ and
$NMC_{\nu}$, respectively, for various values of $M$ and $K$.
When comparing the red curve ($M=1$ and $K=0$) and green curve ($M=10$
and $K=0$) in Fig. \ref{fig7}b, 
one can see that $NM_{\nu}$ is enhanced for
nonlinear tasks in $\tau<5$ as $M$ increases, whereas the memory
capacity decreases in a region of $\tau>5$.
The memory loss is compensated by increasing the number of the past
nodes $K$; thus, $NMC_{\nu}$ can be enhanced when $M$ and $K$ both
increase, as shown by the blue curve ($M=10$ and $K=5$) and pink curve
($M=10$ and $K=10$) in Fig. \ref{fig7}b.
We find that with the time-multiplexing method of $M=10$ and $K=5$, 
$NMC_{\nu}$ for only a single probe $N=1$ can be larger than $NMC_{\nu}$
without the time-multiplexing method (Fig. \ref{fig7}c).
This time-multiplexing method is effective to achieve high RC
performance even when the number of the probes is limited in physical
reservoirs. 

\subsection*{Sensing applications.}
Physical RC frameworks generally suggest that physical systems
responding to input signals {\it themselves} can be utilized as information
processing systems.
This implies that when the physical systems is perturbed by an
external stimulus (e.g., environmental changes),  
the system itself can also be utilized to detect the external stimulus
with an appropriate training process. 
Here, we consider microcavities to detect an environmental 
physical quantity in the RC scheme and demonstrate 
the identification of refractive index $n_{out}$ outside the cavity, i.e., 
refractometric sensing.  

\begin{figure}
\begin{center}
\raisebox{0.0cm}{\includegraphics[width=12cm]{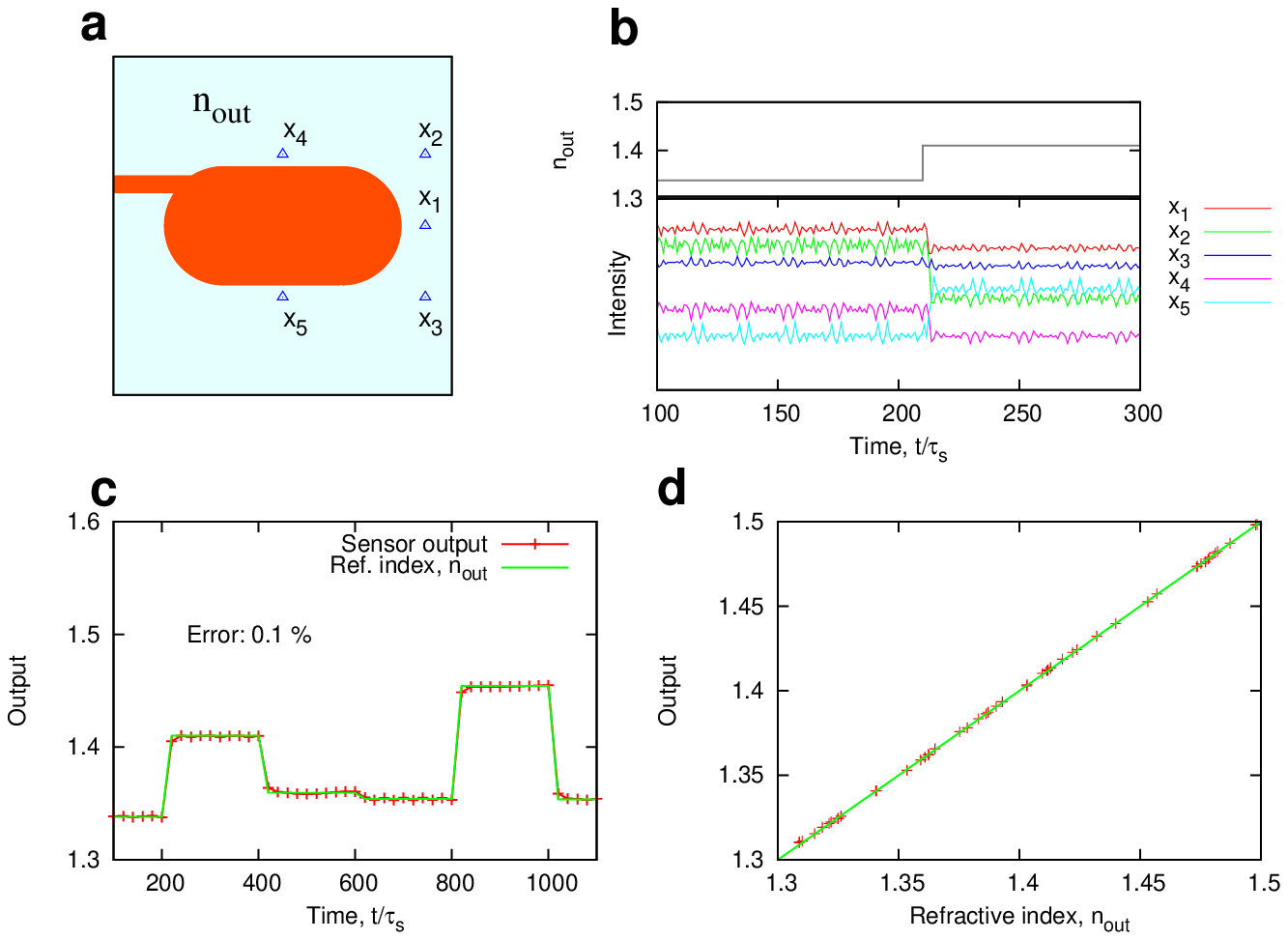}} 
\end{center}
\caption{\label{fig8} 
{\bf Sensing of environmental refractive index}.
{\bf a.} Stadium cavity used for the sensing of refractive index $n_{out}$.
The phase-modulated light is injected into the stadium cavity, and 
five signals detected by probes denoted by the blue triangles are used for 
the output $\hat{y}$.
For passive cavity sensing, $W_{\infty}/W_{th}=0$ was set. 
{\bf b.} An example of the probe signal $x_i$ for the phase-modulated
 light (input). 
$x_i$ changes when $n_{out}$ changes at time $t/\tau_s =215$. 
{\bf c.} The output signal $\hat{y}$ after training. $\hat{y}$ can 
identify $n_{out}$, with 0.1 $\%$ error, even when it rapidly changes. 
{\bf d.} The trained output $\hat{y}$ vs. refractive index $n_{out}$.
}
\end{figure}

As a simple demonstration, we consider the case when a stadium microcavity
is surrounded by a medium with refractive index $n_{out}$, as shown in Fig. \ref{fig8}a.  
A randomly phase-modulated light is injected to the stadium cavity and 
the emission from the cavity are detected with the five probes ($N=5$). 
When the external refractive index $n_{out}$ changes, 
the phases of the reflection/transmission and the coupling to the probes 
are changed; consequently, the detected intensity $x_{i}$ at the probe $i$ 
are also changed (Fig. \ref{fig8}b).  
We use $x_{ij}=x_i(nT_m+j\tau_s)$ ($j=1,\cdots,M$) 
responding to the phase-modulated light to form $\hat{y}=\sum_{i,j}w_{ij}x_{ij}$.
Our purpose is to identify the surrounding refractive index $n_{out}$ from the output 
$\hat{y}$ after the training of $w_{ij}$ to minimize $\|n_{out}-\hat{y}\|$.
In the training process, we used 100 datasets of $\{x_1(t),\cdots,x_N(t)$,$n_{out}\}$, where 
$n_{out}$ was randomly chosen in a region of $1.3 \le n_{out} \le 1.5$.  

Figure \ref{fig8}c shows the trained output $\hat{y}$, where 
it is assumed that $n_{out}$ randomly changes in time.    
Clearly, $\hat{y}$ follows the changes of the index $n_{out}$ 
with the error of 0.1 $\%$, even when $n_{out}$ is rapidly changed in a
timescale of $\tau_s \approx \tau_c$.
Consequently, $n_{out}$ can be identified with low errors, as shown in Fig. \ref{fig8}d. 
We remark that the memory of the reservoir (cavity) does not play an essential role 
in this sensing task.
In this sense, the proposed learning-based sensing scheme is related to 
extreme learning machine (ELM) \cite{Ortin2015} as well as RC.

We emphasize that the proposed method do not need 
any precise sensing model, high-quality microcavity, as well as 
complex post processing,
unlike the previous works on microcavity sensor \cite{Vollmer2012,Hanumegowda2005}, 
where the shift of the resonant frequency in a microcavity 
due to the change of the refractive index has been measured from the
transmission or reflection spectra.  

The results presented in this subsection suggest that 
with the merging of the optical sensing and
learning-based processing, a model-free detection of the external refractive index $n_{out}$ 
is achieved at a rate of $1/\tau_c$. 

\section*{Summary}
In the present work, we proposed and demonstrated RC based on nonlinear wave dynamics 
in a microcavity laser. 
One of the merits of using the wave dynamics is 
to enable high-dimensional mapping of the input information 
into the wave patterns, 
after the nonlinear amplifition in the gain medium;
consequently, a number of the signals detected by each probe can be used as
 nonlinear nodes for high RC performance.
We emphasize that the high-dimensional mapping into the wave patterns 
is spontaneously processed with low energy loss 
by natural multiple reflections, 
at a ultimate short timescale, which may be of the order of a the cavity 
lifetime $\tau_c$, 
suggesting the potential for photonic parallel information processing.   

Then, we also proposed to apply a time-multiplexing encoding technique 
to the wave dynamics and demonstrated the enhancement of the computational capability. 
This method will be useful for the situation when only a few detectors
are available due to a physical constraint, and beyond the example of the microcavity lasers, it will be applicable for any physical RC systems with spatial degree of freedom but only with a few detectors.

Lastly, we discussed the sensing applications of the microcavity-based RC, 
where the microcavity is used as sensing elements as well as a reservoir. 
The combination of the optical sensing and RC will be useful for model-free identification of physical quantities. 

These results opens a way to utilize complex wave dynamical systems at a microscale for fast photonic information processing and will shed light on a novel potential toward model free sensing with the concept of RC. 

\section*{Method}
\subsection*{Maxwell-Bloch model in microcavity lasers}
We assume that the thicknesses of microcavities are much smaller than 
their in-plane dimensions, and 
microcavities were treated as two-dimensional objects by applying
effective refractive indices $n_{in}$. 
To describe the light-matter interaction, we used the Maxwell-Bloch
(MB) model, where the gain medium inside the cavity is modeled as a
two-level system.
The Maxwell-Bloch model is a simple model to
describe the laser dynamics but can quanlitatively reproduce lasing
phenomena in two-dimensional microcavity lasers \cite{Harayama2005,Harayama2003}.
The normalized Maxwell-Bloch model is given by :
\begin{eqnarray}
\dfrac{\partial^2 E}{\partial t^2}
= \dfrac{1}{\epsilon}\nabla^2_{xy}E
-\sigma\dfrac{\partial E}{\partial t}
-\Theta\dfrac{4\pi}{\epsilon}\frac{\partial^2}{\partial t^2}
\left(
\rho + c.c
\right),
\label{MB1}
\end{eqnarray}
\begin{eqnarray}
\dfrac{\partial \rho}{\partial t}
=
-
\left(
\gamma_{\perp}+ 
i\Delta_a
\right)
\rho
+
\gamma_{\perp} 
WE, \label{MB2} 
\end{eqnarray}
\begin{eqnarray}
\dfrac{\partial W}{\partial t}
=
-
\gamma_{\parallel} 
\left(W-W_{\infty}
\right)
- 2iE\gamma_{\parallel}
\left( \rho-\rho^*
\right), \label{MB3}
\end{eqnarray}
where space and time are made dimensionless by the scale transformations
$\omega_sx/c \rightarrow x$ and $\omega_st \rightarrow t$,
respectively. 
$\omega_s$ is a reference frequency close to the transition frequency
$\omega_a$ of the two-level gain medium. 
In Eqs. (\ref{MB1})--(\ref{MB3}), $E$, $\rho$, $W$, and all of the other parameters
are also made dimensionless. 
The normalization is similar to that reported in Ref. \cite{Pick2015}.
$\epsilon =n_r^2$ is the relative permittivity, where the
refractive index $n_r$ is $n_{in}$ inside the cavity and waveguide, whereas 
it is $n=n_{out}$ outside the cavity and waveguide.  
$\sigma$ represents the background absorption inside the cavity.
$\Theta(x,y)$ is a step function; $\Theta$ is 1 inside the cavity and zero
outside the cavity.   
$\Delta_a=\omega_a/\omega_s$ represents the normalized gain center. 
The two relaxation parameters, $\gamma_{\perp}$ and
$\gamma_{\parallel}$, are the
transverse and longitudinal relaxation rates,
respectively. 
$W_{\infty}$ represents the pumping power.  

The Maxwell Eq. (\ref{MB1}) was simulated
by the finite-difference time-domain (FDTD) method, where
a perfect matched layer (PML) was introduced near the boundary of
the calculation space to absorb the emission light. 
See Ref. \cite{Harayama2005} for the simulation method. 

In the stadium cavity shown in Fig. \ref{fig1}, 
the radius $R$ of the half circle and major axis length $L=4R$ 
of the stadium were set $12.25/\sqrt{2}$ and $49/\sqrt{2}$ in a unit of
$c/\omega_0$, where 
$\omega_0$ is the input angular frequency, and $c$ is the light velocity
in vacuum. 
The actual $L$ would be $\sim$1.42 $\mu m$ if $n_{in}=3.3$ and
wavelength $\lambda =0.85 \mu$m.
Although $L$ is shorter than that of a standard microcavity, 
we restrict ourselves to the cases of the short length
due to the lack of the computational power.
We emphasize that the similar results can be essentially obtained in
the cases of the longer lengths. 

The values of the remaining 
parameters are: 
$n_{in}=3.3$, $\sigma=10^{-3}$, $\Delta_a=1$, $\gamma_{\perp}=0.1$,
$\gamma_{\parallel}=10^{-4}$. 
$n_{out}$ was set to be 1 for the results shown in
Figs. \ref{fig1}--\ref{fig7}, whereas  
in Fig. \ref{fig8}, $n_{out}$ is changed in a range of $1.3 \le n_{out}
\le 1.5$. 
We confirmed that these parameter values do not essentially affect the RC performance. 

\subsection*{Incident wave}
The incident light $E_{in}$ is phase-modulated with the input signal $u(n)$, 
and it is injected to the cavity via the input waveguide shown in
Fig. \ref{fig1},
\begin{eqnarray}
E_{in}=A\psi\cos(\omega_0t+\phi(t)), 
\end{eqnarray}
where $A$ is the amplitude, $\psi$ is the lowest-order waveguide mode,
$\omega_0$ is the center frequency of the input light, and it is tuned
to a resonant frequency of the stadium cavity. 
The amplitude $A$ is given such that 
the injection locking to the lasing mode with frequency $\omega_0$ 
is achieved when $W_{\infty}/W_{th}>1$. 
All results presented in this paper are given under the injection
locking condition.

$\psi(t)=\psi_0Mask(t)u(t)$ is the modulated phase, where $\psi_0$ is
the modulation amplitude, $Mask(t)$ and $u(t)$ represents the mask
signal and the input signal at time $t$, respectively. 
$u(t)$ holds for a time interval $\tau_s \approx \tau_c = 143/\omega_0$, 
where $\tau_c$ is the cavity lifetime.
For a small cavity with $R=12.25/\sqrt{2}$ and input wavelength $\lambda=0.85$ nm, 
the cavity lifetime $\tau_c$ is close to 0.1 ps.
Thus, the sampling at the interval $\tau_s\approx \tau_c$ is unrealistic;
however, we remark that the problem can be moderate for a large cavity
because $\tau_c$ can increase with increasing $R$ \cite{Cao2015}.  

As for the mask signal, $Mask(t)$ is a colored noise signal with a 
the decay rate of $1/\tau_c$, which is repeated with a time interval
$T_m$.
The use of such a colored noise can efficiently
excite the modes used for the RC \cite{Kuriki2018}.    

\subsection*{Estimation of parameter values toward actual experiments}
In this study, we have restricted ourselves to the case of the small
cavity of $L\approx 1.67\lambda$ and owing to the limitation of our
computational power. 
However, we remark that the results presented in this paper do not essentially
depends on the cavity size in the normalized form; thus, 
the real values in actual experiments can be estimated.   
For example, when $L=100 \mu$m, average mode interval $\Delta \omega_i/(2\pi)$
will be of the order of gigahertz \cite{Sunada2013}. 
The photon lifetime $\tau_c$ in the stadium cavity (without the
waveguide) is estimated as 28 ps, and it can be changed by the
amplification due to the gain medium.  
Then, the sampling rate $\tau_s$ can be of the order of $\tau_c$, 
which can be set in an actual experiment. 
In addition, we also remark that for a large cavity with $L=100 \mu$m,
the probes or single-mode waveguides coupled to detectors can be relatively
easy to be placed around the cavity.
   
\subsection*{Consistency}
Consistency is the similarity of the response outputs for a repeated drive
signal, and it is one of the important properties of RC \cite{Uchida2004}. 
The consistency can be measured by the correlation between the two
response outputs obtained from different initial conditions.   
We measured the consistency of the wave dynamics in microcavity lasers
driven by the input light $E_{in}$ as follows:
\begin{eqnarray}
C = \dfrac{1}{N}\sum_{i=1}^N C_i,
\end{eqnarray}
where
\begin{eqnarray}
C_i = \dfrac{
\langle (x^{(1)}_i-\bar{x}^{(1)}_i)(x^{(2)}_i-\bar{x}^{(2)}_i) \rangle
}
{
\sigma_{i,1}\sigma_{i,2}
}. 
\end{eqnarray}
$x_i^{(j)}(t)$ denotes the intensity signal detected by the probe $i$,
which is obtained from an initial state labeled by $j$.
$\langle \cdot\rangle$ denotes the time average.
$\bar{x}^{(j)}_i$ and $\sigma_{i,j}$ denote the time average and standard
deviation of $x_i^{(j)}$, respectively. 
By the definition, $C$ is in the range $-1 \le C \le 1$, and it takes
the maximum $C=1$ when the two signals are identical, i.e.,
$x_i^{(1)}(t)=x_i^{(2)}(t)$. 

\section*{Acknowledgments}
This work was in part supported by JSPS KAKENHI Grant
No. 16K04974 and 19H00868, Japan.
S. S. thanks Tomoaki Niiyama for discussions.   
\section*{Author contributions}
S.S. and A.U. conceived the numerical experiments.
S. S. conducted the numerical simulation and analyzed the results.
S.S. mainly wrote the paper, and all authors contributed
to the preparation of the manuscript.
 
\section*{Additional  information}
%
Competing interests: The authors declare no competing interests.

\newpage

\begin{thebibliography}{9}
\bibitem{Versraeten2007}
Verstraeten, D., Schrauwen, B., D'Haene, M., $\&$ Stroobandt, D.
An experimental unification of reservoir computing methods.
{\it Neural. Netw.} {\bf 20}, 391 (2007).

\bibitem{Jaeger2004} 
Jaeger, H.  $\&$ Haas, H. Harnessing nonlinearity: predicting chaotic 
systems and saving energy in wireless communication.
{\it Science} {\bf 304}(5667), 78-80 (2004). 

\bibitem{Maass2002}
Maass, M., Natschlager, T., $\&$ Markram, H.
Real-time computing without stable states: a new framework for neural
computation based on perturbations. 
{\it Neural Comput.} {\bf 14}(11), 2531-2560 (2002). 

\bibitem{Paquot2012}
Paquot, Y. {\it et al.}
Optoelectronic reservoir computing. {\it Sci. Rep.} {\bf 2}(1), 287
	(2012). 

\bibitem{GVderSande2017}
Van der Sande, G., Brunner, D., $\&$ Soriano, M. C.
Advances in photonic reservoir computing.
{\it Nanophotonics} {\bf 6}(3), 561-576 (2017).

\bibitem{Du2017}
Du, C., Cai, F., Zidan, M. A., Ma, W., Lee, S., H. $\&$ Lu, W. D.
Reservoir computing using dynamic memristors for temporal information processing. {\it Nat. Commun.} {\bf 8}, 2204 (2017). 

\bibitem{Nakane2018}
Nakane, R., Tanaka, G., $\&$ Hirose, A.
Reservoir Computing With Spin Waves Excited in a Garnet Film.
{\it IEEE Access} {\bf 6}, 4462-4469 (2018).

\bibitem{Nakajima2015}
Nakajima, K., Hauser, H., Li, T., Pfeifer, R.
Information processing via physical soft body. 
{\it Sci. Rep.} {\bf 5}, 10487 (2015).

\bibitem{Tanaka2019}
Tanaka, G. {\it et al.}
Recent Advances in Physical Reservoir Computing: A Review.
{\it Neural Networks} {\bf 115}, 100-123 (2019).

\bibitem{Brunner2013}
Brunner, D., Soriano, M. C., Mirasso, C. R., $\&$ Fischer, I. 
Parallel photonic information processing at gigabyte per second data rates using transient states. 
{\it Nat. Commun.} {\bf 4}, 1364 (2013). 

\bibitem{Larger2017}
Larger, L. {\it et al.} 
High-Speed Photonic Reservoir Computing Using a Time-Delay-Based
	Architecture: Million Words per Second Classification.
{\it Phys. Rev. X} {\bf 7}, 011015 (2017). 

\bibitem{Vandoorne2014}
Vandoorne, K. {\it et al.}
Experimental demonstration of reservoir computing on a silicon photonics chip. 
{\it Nat. Commun.} {\bf 5}, 3541 (2014). 

\bibitem{Takano2018}
Takano, K. {\it et al.}
Compact reservoir computing with a photonic integrated circuit.
{\it Opt. Express} {\bf 26}(22), 29424-29439 (2018). 

\bibitem{Laporte2018}
Laporte, F., Katumba, A., Dambre, J., $\&$ Bienstman, P.
Numerical demonstration of neuromorphic computing with photonic
crystal cavities.
{\it Opt. Express}  {\bf 26}(7), 7955-7964 (2018).

\bibitem{Sunada2018}
Sunada, S., Arai, K., $\&$ Uchida, A., 
Wave dynamical reservoir computing at a microscale.
Proc. of 2018 International Symposium on
Nonlinear Theory and Its Applications (NOLTA 2018) {\bf 1}, 154-155
(2018).

\bibitem{Dambre2012}
Dambre, J., Verstraeten, D., Schrauwen, B., $\&$ Massar, S.
Information Processing Capacity of Dynamical Systems.
{\it Sci. Rep.} {\bf 2}, 514 (2012).

\bibitem{Appeltant2011}
Appeltant, L. {\it et al.}
Information processing using a single dynamical node as complex system.
{\it Nat. Commun.} {\bf 2}, 468 (2011). 

\bibitem{Martinenghi2012}
Martinenghi, R., Rybalko, S., Jacquot, M., Chembo, Y. K., $\&$ Larger,
L.,  
Photonic nonlinear transient computing with multiple-delay wavelength
dynamics. 
{\it Phys. Rev. Lett.} {\bf 108}(24), 244101 (2012).

\bibitem{Duport2012}
Duport, F., Schneider, B., Smerieri, A., Haelterman, M., $\&$ Massar,
S. 
All-optical reservoir computing. 
{\it Opt.Express} {\bf 20}(20), 22783-22795 (2012).

\bibitem{OpticalMicrocavity1996}
Chang, R. K. $\&$ Campillo, A. L. (eds)
Optical Processes in Microcavities. (World Scientific, New York, 1996). 

\bibitem{NS1997}
N\"ockel, J. U. $\&$  Stone, A. D., 
Ray and wave chaos in asymmetric resonant optical cavities. 
{\it Nature} {\bf 385}, 45-47 (1997).

\bibitem{Cao2015}
Cao, H., $\&$ Wiersig, J.
Dielectric microcavities: Model systems for wave chaos and
	non-Hermitian physics.
{\it Rev. Mod. Phys.} {\bf 87}, 61 (2015).

\bibitem{Harayama2011}
Harayama T., $\&$ Shinohara, S.
Two-dimensional microcavity lasers.
 {\it Laser Photonics Rev.} {\bf 5}, 247 (2011).

\bibitem{Bitter2018}
S. Bittner, S. Guazzotti, Y. Zeng, X. Hu, H. Y?lmaz1, K. Kim, S. S. Oh,
Q. J. Wang, O. Hess, and H. Cao,
``Suppressing spatiotemporal lasing instabilities with wave-chaotic
microcavities,''
Science {\bf 361}(6408), 1225-1231 (2018).

\bibitem{Stadium}
Bunimovich,  L. A. 
On the ergodic properties of nowhere dispersing
billiards.
{\it Commun. Math. Phys.} {\bf 65}(3), 295-312 (1979).

\bibitem{Harayama2005}
Harayama, T., Sunada, S., $\&$ Ikeda, K. S.
Theory of two-dimensional microcavity lasers.
{\it Phys. Rev. A} {\bf 72}, 013803 (2005). 

\bibitem{Harayama2003}
Harayama, T., Fukushima, T., Sunada, S., $\&$ Ikeda, K. S.
Asymmetric Stationary Lasing Patterns in 2D Symmetric Microcavities.
{\it Phys. Rev. Lett.} {\bf 91}, 073903 (2003). 

\bibitem{Sunada2013}
Sunada, S., Fukushima, T., Shinohara, S., $\&$ Harayama, T.
Stable single-wavelength emission from fully chaotic microcavity lasers.
{\it Phys. Rev. A} {\bf 88}, 013802 (2013).

\bibitem{Inubushi2017}
Inubushi M. $\&$ Yoshimura, K.
Reservoir Computing Beyond Memory-Nonlinearity Trade-off.
{\it Sci. Rep.} {\bf 7}(1), 10199 (2017).

\bibitem{Ortin2015}
Ortin, S. {\it et al.} 
A Unified Framework for Reservoir Computing and Extreme Learning
Machines based on a Single Time-delayed Neuron.
{\it Sci. Rep.} {\bf 5} 14945 (2015).

\bibitem{Uchida2004}
Uchida, A., McAllister, R. $\&$ Roy, R. Consistency of Nonlinear System
Response to Complex Drive Signals. 
{\it Phys. Rev. Lett.} {\bf 93}, 244102 (2004). 

\bibitem{Nakayama2016}
Nakayama, J., Kanno, K., $\&$ Uchida, A.,
Laser dynamical reservoir computing with consistency: an approach of a
chaos mask signal.
{\it Opt. Express} {\bf 24}(8), 8679-8692 (2016).

\bibitem{Harayama2015}
Harayama, T., $\&$ Shinohara, S.
Ray-wave correspondence in chaotic dielectric billiards. 
Phys. Rev. E {\bf 92}(4), 042916 (2015).

\bibitem{Takens1981}
Takens, F.
Detecting strange attractors in turbulence. 
In D. A. Rand and L.-S. Young (ed.). {\it Dynamical Systems and Turbulence, 
Lecture Notes in Mathematics}, {\bf 898} Springer-Verlag. 366–381 (1981).

\bibitem{Kuriki2018}
Kuriki, Y., Nakayama, J., Takano, K., $\&$ Uchida, A.
Impact of input mask signals on delay-based photonic reservoir computing with semiconductor lasers.
{\it Opt. Express} {\bf 26}(5), 5777-5788 (2018).

\bibitem{Vollmer2012}
Vollmer F., $\&$ Yang, L.  
Review label-free detection with high-Q microcavities: a review of biosensing mechanisms for integrated devices.
{\it Nanophotonics} {\bf 1}, pp. 267-291 (2012).

\bibitem{Hanumegowda2005}
Hanumegowda, N., Stica, C., Patel, B., White, I., $\&$ Fan, X. 
Refractometric sensors based on microsphere resonators.
{\it Appl. Phys. Lett.} {\bf 87} (2005).

\bibitem{Pick2015}
Pick, A., {\it et al.}
{\it Ab initio} multimode linewidth theory for arbitrary inhomogeneous
laser cavities.
{\it Phys. Rev. A} {\bf 91}, 063806 (2015). 

\bibitem{Shinohara2007}
Shinohara, S. and Harayama, T., 
Signature of ray chaos in quasibound wave functions for a stadium-shaped
dielectric cavity.
{\it Phys. Rev. E} {\bf 75},036216 (2007).
\end{thebibliography}
\end{document}